\documentclass[conference,compsoc]{IEEEtran}
\IEEEoverridecommandlockouts
\usepackage{cite}
\usepackage{amsmath,amssymb,amsfonts}
\usepackage{algorithmic}
\usepackage{graphicx}
\usepackage{textcomp}
\usepackage{xcolor}

\usepackage{tabularx}
\usepackage{longtable}
\usepackage{verbatim}
\usepackage{listings}

\def\BibTeX{{\rm B\kern-.05em{\sc i\kern-.025em b}\kern-.08em
    T\kern-.1667em\lower.7ex\hbox{E}\kern-.125emX}}
\begin{document}

\title{IsoEx - An explainable unsupervised approach to process event logs' cyber investigation}

\author{\IEEEauthorblockN{Pierre Lavieille}
\IEEEauthorblockA{\textit{Unaffiliated} \\
Paris, France \\
pierre.lavieille@yahoo.com}
\and
\IEEEauthorblockN{Ismaïl Alaoui Hassani Atlas}
\IEEEauthorblockA{\textit{Unaffiliated} \\
Paris, France \\
ala-ism@hotmail.fr}
}


\maketitle

\begin{abstract}
39 seconds. That's the timelapse between two consecutive cyber attacks as of 2023. Meaning that by the time you're done reading this abstract, about 1 or 2 additional cyber attacks would have occurred somewhere in the world. In this context of highly increased frequency of cyber threats, Security Operation Centers (SOC) and Computer Emergency Response Teams (CERT) can be overwhelmed. In order to relieve the cybersecurity teams in their investigative effort and help them focus on more added-value tasks, machine learning approaches and methods started to emerge.
This paper introduces a novel method, IsoEx, for detecting anomalous and potentially problematic command lines during the investigation of contaminated devices.
IsoEx is built around a set of features that leverages the log structure of the command line, as well as its parent/child relationship, to achieve a greater accuracy than traditional methods.
To detect anomalies, IsoEx resorts to an unsupervised anomaly detection technique that is both highly sensitive and lightweight.
A key contribution of the paper is its emphasis on interpretability, achieved through the features themselves and the application of eXplainable Artificial Intelligence (XAI) techniques and visualizations.
This is critical to ensure the adoption of the method by SOC and CERT teams, as the paper argues that the current literature on machine learning for log investigation has not adequately addressed the issue of explainability.
This method was proven efficient in a real-life environment as it was built to support a company's SOC and CERT.
\end{abstract}


\section{Introduction}
\begin{figure}[!h]
    \centering
    \includegraphics[width=8cm]{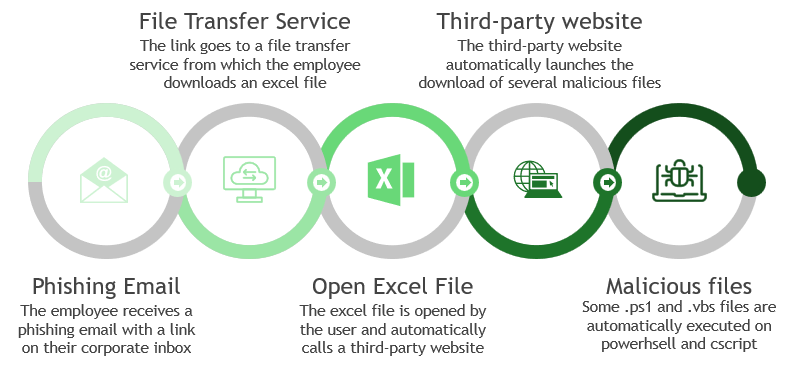}
    \caption{Diagram of a example classic cyber-attack's steps}
    \label{fig:attack_example}
\end{figure}

Cybersecurity Venture predicts that cyber-crime will cost 10.5 trillion USD annually by 2025. With the multiplication of cyber attack tools and media as well as the increasing use of such tools for warfare or political pressure, it comes as no surprise that these amounts of money would be reached by then. And if you top it all off with the fact that cyber attack tools are accessible to an ever wider audience, the cost of cybersecurity might become too much to bear for individuals and corporations alike. \newline
To face this increasing flow of cyber attacks, SOC and CERT teams increase their sizes and spend many hours investigating threats, analyzing their structure, identifying their origin. And that's where the shoe pinches. This investigative effort can become quite the time-consuming task. \newline
An example of a classic cyber-attack can be found in the diagram of Figure \ref{fig:attack_example}. Each step of this attack will produce one or several command-lines which are traces of the cyber attack. The goal of the analytics team of SOC and CERT is to find among the thousands of logs in the device journal, one of those few logs in order to trace the origin of the attack. \newline
This paper is going to focus on how unsupervised learning methods of machine learning could support cybersecurity teams in their investigative effort of a contaminated device's command lines. So that more time can be dedicated to critical tasks where their know-how and expertise would be better spent instead of spending hours on exploring seemingly never-ending log files. For instance, Figure \ref{fig:attack_example} displays an example where the IsoEx methods are able to raise several command lines to the analyst.\newline

To face the increasing amount of data (logs, process events, network events, etc.) cybersecurity teams have to process, the first approaches to emerge were the unsupervised models, such as clustering.  For instance, the Drain model \cite{He.2017} would reduce the dimension of the data the SOC and CERT would have to process in order to identify the malicious logs. Later on, some supervised methods started appearing to try and identify the malicious element directly like Log-BERT \cite{Guo.2021}. This field of study was also tackled by some unsupervised approaches like AnoMark \cite{Junius2022}.\newline

The approach presented in this paper combines two important aspects: the construction of a robust log analysis model that incorporates multiple features for log qualification, and the development of an explainability method that enables cybersecurity teams to interpret the results using familiar concepts commonly used by SOC and CERT teams. These concepts revolve around the parent-child relationship, obfuscation, and other relevant factors. In essence, the IsoEx approach focuses on creating an unsupervised learning framework that is highly explainable and that can be applied to the analysis of process events.\\

As the threat landscape expands, it becomes increasingly important to develop effective techniques for detecting and analyzing anomalous behaviors in command lines. By leveraging the log structure and parent/child relationship of command lines, IsoEx offers a powerful approach to identify potentially problematic activities and detect hidden malware within these devices. The emphasis on interpretability further enables SOC and CERT teams to gain valuable insights into the underlying reasons behind detected anomalies, empowering them to take proactive measures in safeguarding smart peripheral devices from emerging cybersecurity threats. \\

To start with, the paper will present a quick overview of the literature on the topic: this paper would not have been possible without the giants on which shoulders it stands and without all the insights stemming from them.
The second part will present the data set that was used and the features the model uses to evaluate the process events.
The third part will introduce the model and explain all it stages in details. 
The fourth part will present the model's performance and will conclude this paper. \\ 

Due to security concerns and to prevent potential misuse, we have made the decision to not release the source code of our tool. Our primary objective is to safeguard against providing cybercriminals with an opportunity to test their attacks using our tool prior to execution. 
Furthermore, while we have chosen not to release the source code as open source, it is worth noting that the steps and methodologies employed in our tool will be thoroughly explained and documented in this paper. This provides an avenue for interested researchers and practitioners to recreate the code and implement it based on the provided information.

\section{Literature review}

\paragraph{Logs vs. Processes}

In cybersecurity, system logs or logs are widely used and refer to a record of events or activities that have occurred on a computer system, typically maintained by the operating system or applications running on the system. Such log data is universally available in nearly all computer systems and is a valuable resource for understanding system status. \newline
This paper focuses on a specific type of logs: process events. Process events are a specific type of event that is related to the creation, termination, or modification of a process running on the computer. Process events are generated by the operating system or other software running on the computer. In a word, a log is a record of events that includes process events, as well as other types of events that occur on a computer system; while process events are a specific type of event that are related to the creation, termination, or modification of a running process.\newline
Logs and process events are often used in cybersecurity to look out for suspicious or malicious activity on a system. By reviewing logs and processes, analysts can identify patterns of activity that may indicate an ongoing attack or other security incidents. Process events can be especially critical to monitor, as they can provide insights into the behavior of running processes on a system. This in turn can help identify malware or any other malicious activity. Even if the IsoEx method uses process events to detect anomalies, we will also discuss the literature about the detection of malicious activity through log data as logs do incorporate process events data. \\

For the past 35 years the most common method to identify malware from log and process data was rule-based detection. For instance, the first antivirus technologies used ‘signatures', which consist of a cryptographic hash of byte strings, found in certain malicious software, to identify other examples of a given piece of malware \cite{Nachenberg.1997}. All the signatures of malicious malware are called Indicator of Compromise (IOC). Those indicators can also be IP addresses, specific system files or patterns of network traffic that are indicative of malware communication. This IOC method is still used in various companies but it's not always effective against advanced or unknown threats. Other classic defense mechanisms are ‘rule-based’ and can be thought of as a decision tree with hard-coded parameters, often extracted through trial and error and expert knowledge, and taken from real-world examples. \newline
In a word, with more devices, in more places than ever, the old ways of detecting potential security risks fail to keep up with the scale, scope and complexity of the current paradigm. Traditional software systems simply cannot keep up with the sheer number of new malware created every week. And that is exactly where artificial intelligence technologies thrive. This is a great field of application for machine learning algorithms, especially in anomaly detection as it can help hint towards specific and previously unknown situations.  \\

Two types of machine Learning applications in AI for Cybersecurity can be highlighted: \textbf{automated threat detection with machine learning} and \textbf{analyst-led operations assisted by AI}. \newline
In the first use case machine learning enables organizations to automate manual work, especially in processes where it is critical to maintain high levels of accuracy and to respond with machine-level speed – such as automatic threat detection and response, or classifying new adversary patterns. Applying machine learning in these scenarios augments signature-based methods of threat detection with a generalized approach that learns the differences between benign and malicious samples and can rapidly detect new in-the-wild threats. \newline
In the second use case, machine learning models assist in analyst-led investigations by alerting teams to investigate detections or by providing prioritized vulnerabilities for patching. Analyst review can be especially valuable in scenarios where there is insufficient data for models to predict outcomes with high degrees of confidence or to investigate benign-appearing behavior that may go unalerted by malware classifiers. In such a use case the need of Explainable AI is crucial. IsoEx fits into this second category. \\

As logs are unstructured textual data they lend themselves by nature more easily to a Natural Language Process (NLP) analysis. Regarding the supervised NLP techniques, they are often structured in three steps. First, a log parser is used to transform log messages into log keys. Two common log parsing approaches that are commonly used are Drain \cite{He.2017} and Spell \cite{Du.2016}. Then, a feature extraction approach is used to build a feature vector to represent a sequence of log keys in a sliding window. Finally a machine learning or Deep Learning algorithm is used to detect the anomalous sequences.  One of the first models to use deep neural network to detect anomalies was DeepLog \cite{Du.2017}, which does so by using Long Short-Term Memory methods (LSTM) \cite{Hochreiter.1997}. In a similar fashion, LogBERT \cite{Guo.2021} uses Bidirectional Encoder Representations from Transformer (BERT) \cite{Devlin.2019} to detect anomalies among a sequence of log data. The key intuition behind those two architectures is from natural language processing: log entries are viewed as elements of a sequence that follows certain patterns and grammar rules. \newline
In addition to those “classic” NLP techniques, new methods such as few-shot learning that learn anomalous behaviors from very few labeled examples \cite{Zhou.2021} or graph based methods that represent the sequence of logs as a graph and train Gated Graph Neural Network (GGNN) on it (DeepTraLog \cite{Zhang.2022}) seem promising. \newline
Regarding more specific work on Process Events some dedicated methods also exist, such as the method proposed by T. Ongun et al. \cite{Ongun.2021} which selects uncertain and anomalous process events thanks to a Random Forrest Classifier applied on transformed logs with cmd2vec (Mixture of FastText \cite{bojanowski2017enriching} and word2vec \cite{mikolov2013efficient}) command line of parent and child processes. \\

One of the biggest limit of supervised AI models is the size of the data sets it is trained on: Cybersecurity teams must acquire many distinct sets of malware codes, non-malicious codes, and anomalies in order to train a reliant Deep model. Acquiring all of these data sets is time-consuming and requires investments that most organizations cannot afford. Without huge volumes of data and events, supervised machine learning algorithm can provide incorrect results and/or false positives. And getting inaccurate data from unreliable sources can even backfire. Moreover the supervised models assume a ground truth at a point in time, however malware authors have proven adept at finding new ways of exploiting vulnerabilities in cybersecurity software. Therefore supervised methods may be less robust over time, especially to brand new approaches.\newline
Most unsupervised methods with log data rely on the same three key steps as supervised ones. But after the feature extraction phase, they use some unsupervised methods like Principal Component Analysis (PCA) \cite{Xu.2019} or one-class classification models like one-class SVM \cite{Wang.2004}. We also find in the literature unsupervised methods that rely on deep learning, like LogAnomaly \cite{Meng.2019}. These techniques combine template2vec, a method to extract the semantic information hidden in log templates, and Bidirectional LSTM \cite{Schuster.1997}, a deep learning model, to detect anomalies of the whole log sequence. \newline
Another interesting approach has been proposed in a method called Anomark \cite{Junius2022}. It is an unsupervised algorithm which offers a way to train a theoretical model on command lines datasets considered as clean. Once done it is able to detect malicious command lines on other datasets. Anomark uses NLP techniques based on Markov Chains and n-grams. It has the benefit of being efficient, light and easy to train.\newline
Regarding more specific work on process events some dedicated methods also exist, such as the one created by Wang et al. \cite{Wang.2020} which detects abnormal causal paths within system provenance graphs thanks to a customized order-preserving doc2vec for text features and a Local Outlier Factor (LOF) model. 
Another interresting approach regarding process events was the method proposed by Utz Nisslmueller \cite{Nisslmueller.2022} which extracts some features from parent-child process pairs, with a particular focus on the
command line of both, and then uses an anomaly detection algorithm. \newline 
Nonetheless, unsupervised anomaly detection methods are capable of higher sensitivity, especially for detecting novel attacks, a principal reason that they have not been more widely adopted in practice is their inherently higher false positive rate relative to supervised methods \cite{Patcha.2007}. \\

Unfortunately most of the methods we presented to detect anomalies from logs or process events lack transparency. In addition to being extremely costly in resources, neural networks, that are widely used in log investigation, are hardly explainable black box-like methods. Those methods do not provide explanations on how they reached the generated result. Nonetheless, a better understanding of this technology by the analyst and the operators is profitable as it makes for more acceptance and trust in the algorithm's suggestions, more ease of debugging and finally a better assessment of the features by the specialist/users. \newline
Based on this observation, the concept of eXplainable Artificial Intelligence (XAI) emerged in the data science community and is particularly needed in a field like cybersecurity where the specialists need to have a precise understanding of the situation in order to take crucial decisions.  \newline
In a literature review \cite{Mendes.2023}, the authors investigate the current state of the art  regarding XAI applied to cybersecurity, aiming to discover which areas of cybersecurity have already benefited from this technology. The paper lists a total of 35 papers which used classic XAI methods like Shapley Additive exPlanations (SHAP) \cite{Lundberg.2017}, Local Interpretable Model-Agnostic Explanations (LIME) \cite{Ribeiro.2016} or techniques like Local Explanation Method using Nonlinear Approximation (LEMNA) \cite{Guo.2018}. Nonetheless, most of this research focuses on Intrusion Detection, Malware Classification and Phishing Detection areas. Many cybersecurity fields do not have any literature on the subject of XAI yet, especially Malware Detection from log data or process events data. This paper means to remedy to that by proposing a brand new method that relies on quantitative features to help the analysts identify anomalous malware. \newline

\section{Dataset structure presentation}
IsoEx was built around proprietary data (obtained through Microsoft Defender’s process events table). The bulk of the gathered data consists of various types of logs. So, before delving deeper into the dataset's structure, let's take a step back and talk a bit about what this paper means by logs and process events.\newline

A log would usually look like the following: 
\begin{center}
2021-04-05T13:24:47.5679556Z identity\_helper.exe --type=utility --lang=en-US --service-sandbox-type=none --mojo-platform-channel-handle=5488 --field-trial-handle=2052 /prefetch:8
\end{center}

Let's break it down and identify the different components of that log

\begin{itemize}
\item 2021-04-05T13:24:47.5679556Z: this is the log's timestamp. The time at which it was generated. 

\item identity\_helper.exe: This is the name of the executable file that's being run.

\item --type=utility: This is a command-line option that specifies the type of program that identity\_helper.exe is running as. 

\item --utility-sub-type=winrt\_app\_id.mojom.WinrtAppIdService: \newline This is another command-line option that specifies the sub-type of utility that identity\_helper.exe is being run as. Specifically, it's being run as a Windows Runtime (WinRT) App ID service.y\_helper.exe is running as. 

\item --lang=en-US: This command-line option specifies the language that the utility will use.y\_helper.exe is running as. 

\item  --service-sandbox-type=none: This command-line option specifies the type of sandboxing that will be used for the utility. In this case, the sandbox type is set to none.y\_helper.exe is running as.

\item --mojo-platform-channel-handle=5488: This is a command-line option that specifies the handle for the Mojo platform channel that the utility will use.y\_helper.exe is running as. 

\item --field-trial-handle=2052,i ...: This command-line option specifies the field trial handle that will be used for the utility.y\_helper.exe is running as. 

\item /prefetch:8: This is a Windows-specific command-line option that specifies how many application files should be preloaded into memory for faster startup. In this case, it's set to preload 8 files.y\_helper.exe is running as. 

\end{itemize}

The data that was gathered through Microsoft Defender contains unlabeled process creation events. The used dataset also presents for each process event the parent-child relationship, some logon information, and the initiating process’ internal filename among others. A detailed view of the table’s headers can be found in the documentation of Microsoft Defender. \newline
It is relevant to note at this point that even though IsoEx was built and tested on this Microsoft Defender generated dataset, the model does work with any other data source, even raw logs. Using raw logs would in fact only require some pre-processing of the input data to make it consumable by IsoEx. Hereafter is a small illustration of how that pre-processing would look like. To sum it up, even though the model was primarily built with Microsoft Defender's datasets, it can still manage to deal with data coming from different EDRs and even with manual uploads of .csv files for instance as long as some basics columns such as the generation timestamp or the parent command line are provided in the data set.\\

\begin{figure}[!h]
    \centering
    \includegraphics[width=8cm]{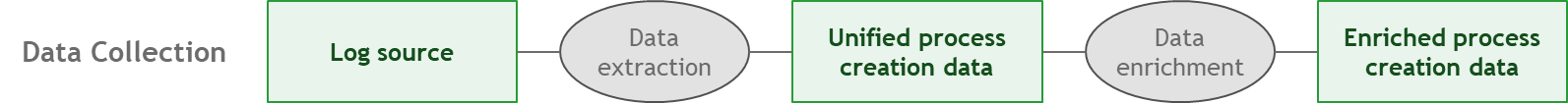}
    \caption{Pre-processing steps to generate an IsoEx-ready dataset from raw logs}
    \label{fig:pre-processing}
\end{figure}

In order to enrich this database and increase the model's accuracy and reliability, the process events table was joined with another table from Microsoft Defender: the DeviceImageLoadEvents table. This table makes it possible to factor in binary files like .dlls when performing the anomaly analysis of logs. A detailed view of the table’s headers can be found in the documentation of Microsoft Defender.

This dataset captures process events and image load events coming from more than 50 devices (client or server) with different types of users (administrator or not) so that a large enough span of scenarios and potential use cases can be represented in the data. On all devices, process events and load image events were captured for a period of 30 days on average (time limit set by Microsoft Defender).

\section{IsoEx Approach}

The goal of IsoEx is to help the analytics teams speed up the detection of malware or other malicious activity thanks to a light unsupervised anomaly detection model trained on process events data. \newline
As log data is unstructured, it was decided to use a mix of rule based and machine learning based approaches to extract features. The choice leaned towards an unsupervised model as the goal is to create an agile solution which would require minimum computational or data resources. Even if unsupervised models are more sensitive and return more false positive we do not perceive this as a hindrance, as this tool will be used by analysts. \newline
Moreover, this paper aims at remedying the lack of XAI in malware detection from logs and processed data. To rely on the prediction made by a machine learning model and act from it, the analyst-led operations team  needs “understanding”. We introduce a brand new method that relies on rule-based features and explainability techniques to help the analysts identify anomalous malware. By resorting to an XAI approach, we anticipate that the IsoEx method will be more readily adopted and will continue to evolve over time through user feedback. Such feedback may ultimately help to reduce the prevalence of false positives associated with unsupervised methods. \newline

\subsection{Feature engineering}
From raw logs or process events data, several approaches are possible for feature extraction. One of them is the rule-based detection approach, where some rules to create variables were defined. These rule-based features are well-suited for detecting subtle signals and have the added benefit of being easily understandable by domain experts. Another solution is the natural language processing approach that was mentioned above when talking about DeepLog or LogBert. We decided not to choose between those two approaches and to create a dataset that is a combination of business rules and machine learning features. We believed that this combination would not only make IsoEx interpretable but also efficient and reliable in the results it provides. Regarding the business rules, we created them thanks to literature and the feedbacks from the expert team we worked with.  \newline

One major challenge of using machine learning in cybersecurity is the potential for cyber-criminals to leverage AI to analyze their malware and develop more advanced attacks. As a result, this paper acknowledges the importance of maintaining ambiguity around the specific details of the created features. Rather than providing a step-by-step guide for creating each feature, this paper presents a set of feature ideas that draw on both cybersecurity rules and machine learning techniques. \newline
From each feature idea, one or multiple features can be created. To prioritize a particular principle, the paper proposes creating multiple features from a single idea. For instance, if a rule generates a number based on the frequency of a specific character in a command line, the paper suggests creating a primary feature using this number and a secondary Boolean feature that maps the top 5\% based on the highest value of this feature. \newline

Below is an implicit list of the features we are considering, and it should be noted that these feature ideas can be applied to either the parent process, child process, or both:

\begin{enumerate}

\item \textbf{IOC and Signature: } 
similarly to traditional methods, detecting anomalies through signature detection and other indicators of compromise (IOCs) can be relevant.

\item \textbf{Obfuscation attempts: } 
we conduct a check for potential obfuscation within the command line string by employing various techniques. One approach involves identifying specific substrings that are commonly associated with obfuscation, as documented in the literature. Another technique involves checking whether the filename is mentioned in the command line, as its absence may indicate potential obfuscation.

\item \textbf{Parameter analysis: }
for each process event, we parse the command line to extract all its parameters and identify any anomalous parameter it may contain. The definition of what constitutes an abnormal parameter can be informed by public documentation and frequency of use. 

\item \textbf{Parent-child Analysis: }
we contend that analyzing the relationship between parent and child processes is critical in identifying outliers among typical events. By scrutinizing the distribution of parent processes for a particular child process and vice versa, we can discern which parent or child processes are abnormal for a given process event. 

\item \textbf{Siblings-child Analysis: }
in addition to analyzing parent-child relationships, it is also important to consider the relationships between siblings and the number of child processes spawned by a given process. A single process can spawn multiple child processes either in parallel or sequentially, and by tracking these events in the process journal, we can count the number of parallel child processes (siblings) and the total number of distinct child processes spawned. Additionally, it is important to account for situations where child processes may end before their parent process, and we propose creating an indicator to measure this aspect as well.

\item \textbf{Documentation: }
we established a database of .exe files deemed safe, which incorporates various online sources and the specialized knowledge of the SOC and CERT team we worked with. This database is regularly updated and enables us to verify whether the use of a particular file is safe and documented. Additionally, our documentation includes the safe locations for each executable, allowing us to verify that a known executable is located in a typical location.

\item \textbf{Presence of Scheduled processes: }
the detection of scheduled processes is another feature we consider. We search for specific substrings or keywords in the command line that are known to be associated with scheduled viruses. Additionally, we explore the use of timestamp analysis as another method to identify scheduled processes.

\item \textbf{Presence of Hex: }
this feature aims to detect the presence of hexadecimal values in command lines by utilizing specific substrings identified through literature or regular expression matching.

\item \textbf{Presence of Base-64: }
the presence of Base-64 encoding in a command line string can be indicative of malicious activity. This feature involves identifying specific substrings through prior research or applying Base-64 expression to detect the occurrence of Base-64 encoding.

\item \textbf{Presence of Phishing: }
this feature aims to detect obvious signs of phishing in the command line by using a list of keywords. One classic example of such a keyword is the double extension (.pdf.exe, .xlsx.exe, etc.) which is a common sign of phishing.

\item \textbf{Extension analysis: }
we check with this variable if the command line contains any suspicious extension. To do this, we compiled a list of known suspicious extensions based on previous research and feedback from the experts we had the opportunity to work with.

\item \textbf{Hostname analysis: }
this feature is used to identify if a process event contains any unusual hostname or website. To achieve this, a list of known suspicious hostnames and websites is created based on literature and input from experts, which is then used for comparison against the hostname present in the process event.

\item \textbf{URL encoded: }
URL encoding is a method for representing certain characters in a URL by replacing them with one or more character sequences. This feature involves detecting such URL-encoded substrings in the command line, either by identifying specific substrings through prior research or using regular expressions designed to detect such encoding in command line.

\item \textbf{Webremote download: }
this feature aims to identify potential download, upload, or command-and-control (C2) behavior through the analysis of command lines. This is done using a list of keywords and indicators related to webremote activity or IP addresses and domains associated with C2 servers. 

\item \textbf{Entropy: }
One unexpected clue for detecting encryption in command lines is its entropy. Safe command lines typically have an entropy range similar to that of English text, while encrypted commands have a significantly higher entropy. If a command line's entropy exceeds a certain threshold, it could be an indication of encryption.

\item \textbf{Dump: }
The "Dump" feature is used to scan the command line for any memory or credential dumps generated by a process. This is done by checking for the presence of certain suspicious keywords in the command line string. The list of keywords used for this purpose is defined based on insights from literature as well as the feedback of our experts.

\item \textbf{Filesystem Paths Analysis: }
It involves verifying if the command line string contains multiple filesystem paths, which may suggest behavior related to copying or alternate data streams (ADS).

\item \textbf{Integrity analysis: }
On Windows-based systems six distinct integrity levels can be denoted: 'Untrusted', 'Low', 'Medium', 'High', 'System' and 'Installer'. Processes that are created by users typically run in the Untrusted - High integrity range, whereas created services tend to run with System privileges. By factoring in these integrity levels in the model one can detect potential anomalous behaviour. \newline
Another interesting aspect with integrity level is to compare the integrity level of a command line with the integrity level of the processes from the same executable. Those potential deviations can be helpful to detect User Account Control bypasses.

\item \textbf{Path length analysis: }
In addition to verifying if the process is launched from the correct path and folder (documentation), analyzing the length of the path and the difference in length between the parent and the child can be useful. The length is determined by counting the number of folders from the system path. It's important to keep in mind that these rules should be considered in relation to the type of executable to identify any unusual pattern.

\item \textbf{Launch time: }
by utilizing the timestamp of a process event, we are able to pinpoint the exact time when a command line was executed. This feature allows us to detect whether a process event occurred at an unusual time, such as during the night or over the weekend.

\item \textbf{Frequent process: }
this feature aims at determining how rarely a common line is used. To do so it can either count the total number of occurrences of a feature, the number of hours or days it is used or a combination of both.

\item \textbf{Execution time: }
by using the duration of the the process we can determine if a process is too slow or too fast compared to similar processes. 

\item \textbf{Time outliers: }
with this variable we model the number of processes realized through time with classic time series methods. From this modeling we can identify outliers, i.e. the time when too many process events occurred compared to the projection of the model.

\item \textbf{Language feature extraction: }
to extract language information from the command line, we can utilize the feature extraction process used in supervised and unsupervised log anomaly detection methods. This involves creating a vector representation of the command line, which can be used to create various features based on the dimensions of the vector. 

\item \textbf{Character distribution: }
for this feature we model the character distribution between lower, upper, number and special characters for each kind of executable file. Then we check for each command line if its character distribution is an outlier regarding its executable file.

\item \textbf{AnoMark: }
as mentioned earlier, Anomark offers a way to train a theoretical model on command line datasets. Once done it is able to detect malicious command lines on other datasets. It can be trained upstream and then applied to the logs data. But it can also be trained and infer on the processes of one device, however it theoretically needs to be trained on "clean" data.

\item \textbf{Image Events: }
image events can be cross-checked with the process event journal to identify anomalous processes that cannot be detected through process events alone. This is done by analyzing the generated images, such as .dll files, and creating features to detect divergent image events similar to those created for executables. The model can use several features, which are similar to what was done with the process events, in order to detect those divergent created image events.

\end{enumerate}

\subsection{Machine Learning model}

Once we obtain our tabular dataset composed of coherent and relevant features, we can advance to the anomaly detection phase. The idea is to create an anomaly score for each process event, in order to rank all the processes of a device from the less abnormal to the most abnormal event. As mentioned earlier, we opted for an unsupervised model as we want to create an agile solution which does not need a lot of resources. \newline 
To accomplish this, we have compiled a list of advanced unsupervised anomaly detection models. We aim for our model to be both efficient and fast, while also being able to provide interpretable results. To assess each model's efficiency, we measure its performance on a set of sample data provided by a SOC and CERT team with whom we collaborated. Additionally, we assess each model's speed and interpretability using XAI methods.
\newline
Here is a summary of all the models we compared: 

\begin{itemize}
\item \textbf{Isolation Forest \cite{Liu.2008}:}
This is a tree-based model that ‘isolates’ observations by randomly selecting a feature and then randomly selecting a split value between the maximum and minimum values of the selected feature. Since recursive partitioning can be represented by a tree structure, the number of splittings required to isolate a sample is equivalent to the path length from the root node to the terminating node. This path length, averaged over a forest of such random trees, is a measure of normality and our decision function. Random partitioning produces noticeably shorter paths for anomalies. Hence, when a forest of random trees collectively produces shorter path lengths for particular samples, they are highly likely to be anomalies.

\item \textbf{Gaussian Mixture Model \cite{Reynolds.2009}:}
This model describes a dataset as a collection of Gaussian distributions, each with its own mean and variance. This is useful because it allows to represent complex patterns in the data as a combination of simpler patterns. To create a Gaussian Mixture Model, we start with an initial guess for the number of Gaussian distributions that might be present in the data, and their parameters (i.e., mean and variance). We then use an algorithm called Expectation-Maximization to update these parameters based on the data. Once we have a Gaussian Mixture Model that describes the data well, we can use it to identify anomalies in the dataset. Anomalies are data points that are very unlikely to have been generated by any of the Gaussian distributions in the model. We can detect anomalies by calculating a score for each data point that measures how well it fits the Gaussian Mixture Model. Data points with low scores are more likely to be anomalies.

\item \textbf{One Class SVM \cite{Schölkopf.1999}:}
One-Class Support Vector Machine is an unsupervised model for anomaly or outlier detection. Unlike the regular supervised SVM, the one-class SVM does not have target labels for the model training process. Instead, it learns the boundary for the normal data points and identifies the data outside the border to be anomalies.

\item \textbf{Elliptic Envelope \cite{Rousseeuw.1999}:}
This model creates an elliptical area around a given dataset. Values that fall inside the envelope are considered normal data and anything outside the envelope is returned as outliers. The dataset needs to be Gaussian distributed.

\item \textbf{Auto Encoder \cite{Bank.2021}:}
This Neural Network model learns two functions: an encoding function that transforms the input data, and a decoding function that recreates the input data from the encoded representation. The autoencoder learns an efficient representation (encoding) for a set of data, typically for dimensionality reduction. The idea of using an auto-encoder for unsupervised anomaly detection is to use the reconstruction loss as an anomaly score. If the reconstruction score is high, the observation is more likely to be an outlier. In this case we used Mean Square Error as the reconstruction loss.

\item \textbf{Local Outlier Factor \cite{Wang.2011}:}
This algorithm is an unsupervised anomaly detection method which computes the local density deviation of a given data point with respect to its neighbors. It considers as outliers the samples that have a substantially lower density than their neighbors. From the density of all the features it creates an anomaly score for each observation.
\end{itemize}

We chose the Isolation Forest model as it met all our requirements. We evaluated the performance of various models on a batch of examples provided by a team of experts. The distribution of the anomaly score was a key factor in our evaluation, as we were looking for weak signals that could be clearly identified. Figure \ref{fig:model_comparison} shows a comparison of the anomaly scores for a single device, where the red line represents the anomaly score of the identified malware. \newline
The Isolation Forest and Gaussian Mixture Model performed the best on these criteria, but the Isolation Forest was faster to train and more interpretable due to its tree-based nature. In fact, the Isolation Forest model was 10 to 12 times faster than the GMM model when calculating interpretability using SHAP.

\subsection{Interpretability}

Once the model trained, we need to make sure to produce insights for the end-users (i.e. SOC and CERT analysts) as to how the model is behaving and why it is producing the results it displays. \newline
In a paper about XAI in Fraud Detection \cite{Psychoula.2021}, the authors compared several supervised and unsupervised models and two explanability methods : SHAP and LIME. It concludes that LIME is Faster but SHAP gives more reliable explanations. Although real-time performance is a crucial aspect of our solution, we prioritize providing trustworthy explanations as they are essential for users to have confidence in our solution. This statement leads us to choose SHAP rather than LIME. \newline
As mentioned earlier, we chose the Isolation Forest model for our analysis because of its tree-based structure. While there are various SHAP explainers available, some of them are model-agnostic explainers, such as the permutation explainer which we used for all models except for Isolation Forest. This explainer obtains exact Shapley values for any model by iterating through an entire permutation of features in both forward and reverse directions. However, this permutation process can be lengthy, especially for datasets with many features like ours. In contrast, Tree SHAP is a faster and exact method for estimating Shapley values for tree models and ensembles of trees. It leverages the tree structure of the models to speed up the calculation of Shapley values. \newline
For each process event, we use SHAP Tree-based models to obtain the exact Shapley value associated with each feature. This enables us to rank the effect of each feature we created on its final anomaly score. By presenting the variables that contributed the most (in absolute value) to the final score of each process event, we can identify the issue the rule points out. For instance, if the untrusted hostname feature is activated and supports a high anomaly score, we can forward said hostname to the user so that they can immediately understand the source of risk. We can also obtain the global importance of each feature on the model's overall results. This allows us to better understand the model and adapt the list of features by reinforcing or removing those with low impact and questioning those with high impact. \\

Beyond those Explainable Artificial Intelligence tools we also use some other "interpretability" techniques to ease the analysis of the operator when they have to deal a large quantity of logs. \newline
The first one is the use of a log parser algorithm to create cluster of logs. This is because groupings of similar logs tend to have similar anomaly scores, which can make their analysis more difficult. By clustering similar events, we limit this issue and provide operators with the ability to focus on homogeneous groups of events, rather than analyzing each process event one by one. We used the DRAIN algorithm to group the process events. However, it is important to note that this method has limitations, as some outliers may be hidden within a cluster of safe logs. \newline
The second approach is the development of a web application to simplify the utilization of our method and its features by analysts. Once an analyst identifies a device that may have issues, they only need to input the device number into the application, and the solution will handle the rest. The application will then present a table of all the logs, ordered by their anomaly score. By selecting a suspicious log, the application will display an interpretability window that uses the associated features and SHAP values to provide insight into the reasons behind the model's abnormal score. Additionally, the application includes various filtering options to enhance the search capabilities of the analyst team. \newline
Finally we also integrate several visualization to enrich the the data presented in the application to facilitate the analyses. Among those plots we can mention the process trees inspired by the Microsoft Threat Intelligence Python Security Tools \cite{HellenMSTICPy.2021}. Once the agent has selected a log to investigate, in addition to the abnormality score and the explanability tools, this process trees allow him/her to envision in a very clear way all the parent processes, the children processes and all sibling processes launched by the same parent and its respective childrens (Figure \ref{fig:mstcipy_example}).

\subsection{Training the model and rare features' approach}

The IsoEx pipeline is designed to be trained and applied on all process events of a single device, typically over the last 30 days. Our approach is centered on the device level, as we have created a tool to assist analysts in detecting anomalies from process events. This is an unsupervised and flexible method which has to be retrained for each device to best suit its properties. However, this approach can lead to some issues when training on a single device. In this section, we will discuss these issues and propose solutions that we have implemented in our use case. \\

The IsoEx pipeline is designed to model only the last 30 days of process events, which can be problematic for rare events. For example, if a device user only uses a program like Excel once in this period, certain features like anomalous parent feature may become less relevant due to the small sample size. However, these features remain important in the general case and cannot be simply deleted. The use of such device-centric modeling could lead to the detection of some unknown malicious malware that is never used by the device. To mitigate this issue, three possible solutions are proposed and can be combined. \newline
The first solution is to create a larger training set that includes process events from similar devices. This would increase the pool of data and avoid the rare events, but may require careful selection of devices to prevent the model from being too general. The second solution, called the "scarcity premium," would always activate some features that depend on the total number of observations of a program, boosting rare events. However, this solution raises the question of what defines a rare event. Finally, the last solution is to use rules that can identify and assign more weight to rare events by creating several variables from these rules. We will use a combination of those 3 solutions. \\

Another interesting issue is the influence of each variable on the final anomaly score. We questioned the importance of the activation rate of a binary variables in the final result. \newline 
Some of our variables are binary variables, either zero if everything is normal or 1 if one of the cyber/data rules we created has been activated (download marker, encoded...). But does every variable weigh the same in the final anomaly score based on the number of times this rule appears among the processes?
For instance, let's suppose two uncorrelated rules A and B, the rule A is activated on 5\% on the processes and the rule B is activated in 1\% of the processes, which rule has the most impact on the final anomaly score?  \newline
According to our observations, rule A will have more impact on average than rule B, as we observed a statistical dependence, between the number of activations of a binary feature and the mean shapley score of a feature. We also discover that the maximum shapley impact of a binary variable is impacted by the total number of times it is activated. \newline
This phenomenon presents potential challenges in various scenarios. When a critical feature is globally rare on a device but serves as a strong indicator of malware presence, its infrequent activation may lead to a lower overall anomaly score. This situation poses a dilemma in anomaly detection, as a rare feature's significance in identifying malicious behavior might not be adequately represented in the model's output.
A classic machine learning approach could be Variance Thresholding, where it is common to drop the feature with a small variance. Nonetheless, we believe this approached is biased for anomaly detection, and that another solution should be adopted.\newline
To overcome this problem we turned to the Synthetic Data Generation. First, we determine a percent threshold of activation for binary features, in a second step we create some synthetic data to ensure that all the binary features are above this threshold in percent, then we train the model on the combination of real and synthetic data and finally we used this created anomaly detection model to add an anomaly score to each process event of the device. Regarding the creation of synthetic data we tried a "controlled" random method and some more synthetic data generation methods \cite{xu2019modeling, kamthe2021copula, arjovsky2017wasserstein, li2020sync}. The key aspect in data generation is to make sure the other features stay coherent. At the end of the day we used CopulaGAN as it was the best compromise between data coherence and calculation time.\\

\begin{figure}[!h]
    \centering
    \includegraphics[width=7cm]{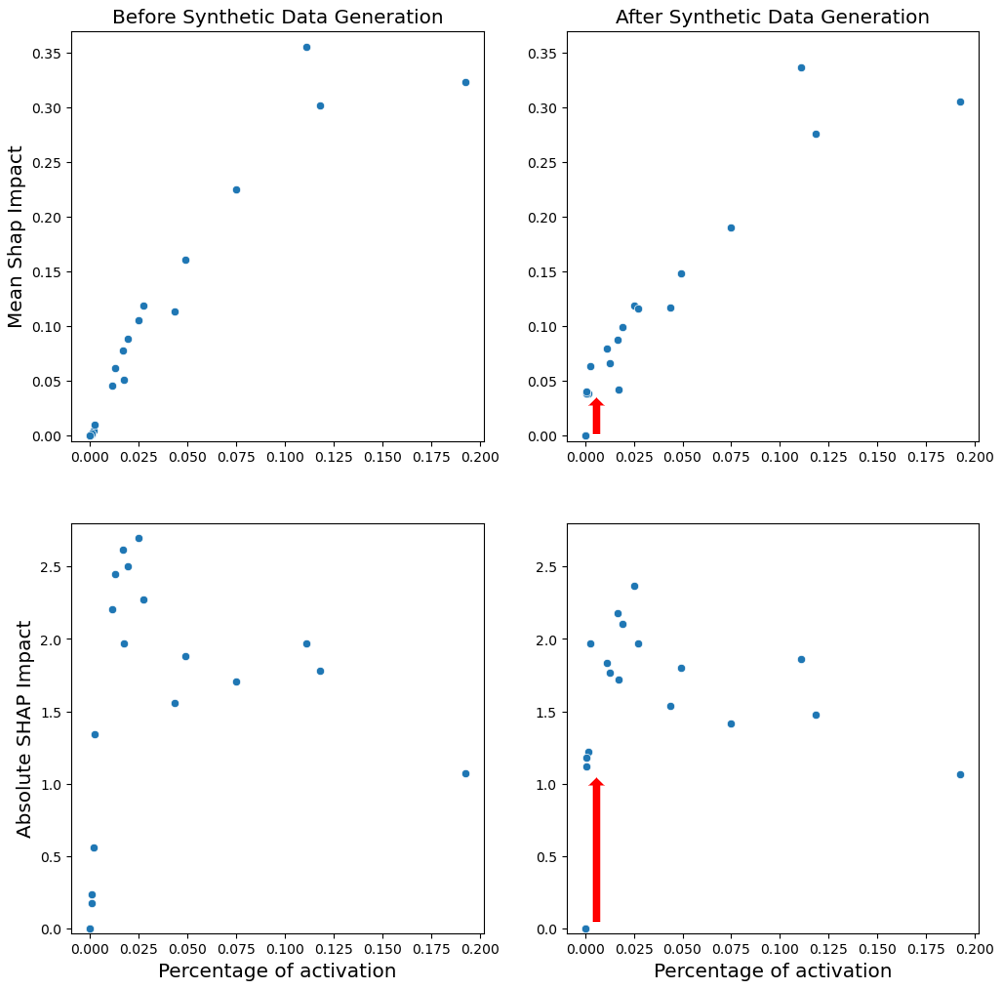}
    \caption{Comparison of the mean and absolute shap effect before and after the use of synthetic data generation for rare features}
    \label{fig:synthetic_effect}
\end{figure}

On Figure \ref{fig:synthetic_effect}, we can observe the effect of our method on the data of a selected device. Each point represents one of the boolean/binary feature of our dataset. For this example we set the threshold at 1\% of activation. On the top left plot we can observe the clear correlation between the percentage of activation and the mean shap effect. When we compare the plot with or without the use of synthetic data, we can immediately notice a big difference for the feature with the smallest percentage of activation. Both their average shap impact and their absolute shap impact makes a jump and brings them closer to the feature with more than 1\% of activation. \newline
Regarding counter effects of this method, we can notice that even the global mean shap impact is stable between the two models, the absolute shap effect can decrease for some features when using the model trained with synthetic data. 

\section{IsoEx results}
The evaluation of the IsoEx pipeline's performance should take into consideration various factors. On the one hand, its ability to detect process events related to malware on a device is critical. On the other hand, its relevance, efficiency in speeding up the work of cybersecurity teams, and interpretability should also be assessed, as the tool was designed to assist cybersecurity teams in their investigation of contaminated device process events.

To evaluate the pipeline's performance, we tested it in real conditions to support SOC and CERT teams. These teams used the tool on a regular basis to assist them in identifying abnormal logs in contaminated devices. This allowed us to test the method on a diverse range of threat types, and each time the model was successful in identifying the anomalous process events in the problem. In fact, the pipeline was even capable of detecting command lines from various stages of a cyber attack, as depicted in Figure \ref{fig:attack_example}. \\

Regrettably, due to confidentiality constraints, we are unable to disclose details about the entirety of the cases addressed using the IsoEx method. Nevertheless, in order to provide some insight into the achieved outcomes, we present the results corresponding to Figure \ref{fig:attack_example} in Table \ref{fig:ResultTablexample}. To maintain data privacy and prevent any potential data leakage, the logs have been partially anonymized. In the table, placeholders indicated by "\textless  \textgreater" represent anonymized information, with varying levels of detail provided.

\begin{table}
\centering
\begin{tabular}{|p{2.5cm}|p{2.5cm}|r|}
\hline
\textbf{Process Event} & \textbf{Parent Process Event} & \textbf{Anomaly Score} \\
\hline
\hline
"msedge.exe" --single-argument \textless link to third party website\textgreater & "EXCEL.EXE" \textless to excel phishing file\textgreater & 66.0 \\
\hline
"protocolhandler.exe" \textless *\textgreater & "msedge.exe" --no-startup-window --win-session-start /prefetch:5 & 63.4 \\
\hline
"SCToastNotification.exe" "Installation complete" \textless *\textgreater & "SCNotification.exe" & 63.3 \\
\hline
fromappreceiver & "UpdaterService.exe" & 63.0 \\
\hline
"msedge.exe" --single-argument \textless external sharepoint link\textgreater & "OUTLOOK.EXE" & 63.0 \\
\hline
"msedge.exe" --single-argument \textless phishing link to wetransfer.com\textgreater & "OUTLOOK.EXE" & 62.7 \\
\hline
"protocolhandler.exe" \textless *\textgreater & "msedge.exe" --no-startup-window --win-session-start /prefetch:5. & 62.2 \\
\hline
"msedge.exe" --single-argument \textless internal sharepoint link\textgreater & "OUTLOOK.EXE" & 62.2 \\
\hline
fromappreceiver & "UpdaterService.exe" & 62.1 \\
\hline
"SCToastNotification.exe" "Downloading and installing software" \textless *\textgreater & "SCNotification.exe" & 62.0 \\
\hline
\end{tabular}
\caption{ Overview of the top 10 process events wound up by IsoEx on an attacked device }
\label{fig:ResultTablexample}
\end{table}

The table presents the top 10 process events ranked by their anomaly scores, as determined by our IsoEx method. It is important to note that the model was trained solely on the process events from the past 30 days of one device, which encompass a total of 17,007 entries. In the table, we provide anonymized process events along with their corresponding parent process events and the associated anomaly scores assigned by IsoEx. The minimum anomaly score recorded for this particular device is 33.7. \newline
Our initial observation highlights a notable distinction between the first and second entries in terms of their anomaly scores, with a gap of 2.6 points. Notably, the first entry corresponds to the fourth step depicted in Figure \ref{fig:attack_example}. Further analysis of Isoex result reveals that several factors contribute to the high anomaly score assigned to this process event, including an anomalous parent process (Excel calling MS Edge), the presence of an untrusted hostname, and indicators of file download. \newline
Additionally, it is noteworthy that multiple processes within the top 10 list exhibit file downloads through Microsoft Edge from Outlook. One such instance is captured in the second step illustrated in Figure \ref{fig:attack_example}, it is the sixth entry in the table involving an Outlook email redirecting to wetransfer.com. The remaining instances involve downloads from SharePoint. In all three cases, the primary indicators of concern are the presence of a download command in the command line and the involvement of an untrusted hostname (even local SharePoint is deemed risky based on input from collaborating analysts). \newline
Concerning the other highlighted process events within the top 10, they are classified as anomalous due to either limited documentation or elevated scores assigned by Anomark, despite their harmless nature.

\section *{Conclusion}
This paper presents a new approach of machine learning to process events' analysis using a combination of unsupervised learning models and explainability modules. \newline
The results are two-fold and show both an overall encouraging performance of the model on different data sets and environments on the one hand and a business-compliant approach on the other as the various SOC and CERT teams that have had the opportunity to try the models were overall satisfied.
The next steps that could be envisioned could be for instance to explore the possibility to reduce the model's sensitivity to anomalous processes that are not necessarily malicious.

\section*{Acknowledgments}

We wish to thank eleven strategy and Erwan Le Pennec PhD for the help they provided for this research.

\newpage
\bibliographystyle{ieeetr}
\bibliography{bibliography-isoex}


\onecolumn
\newpage
\appendix
\section{Appendices}

\begin{figure}[!h]
    \centering
    \includegraphics[width=10cm]{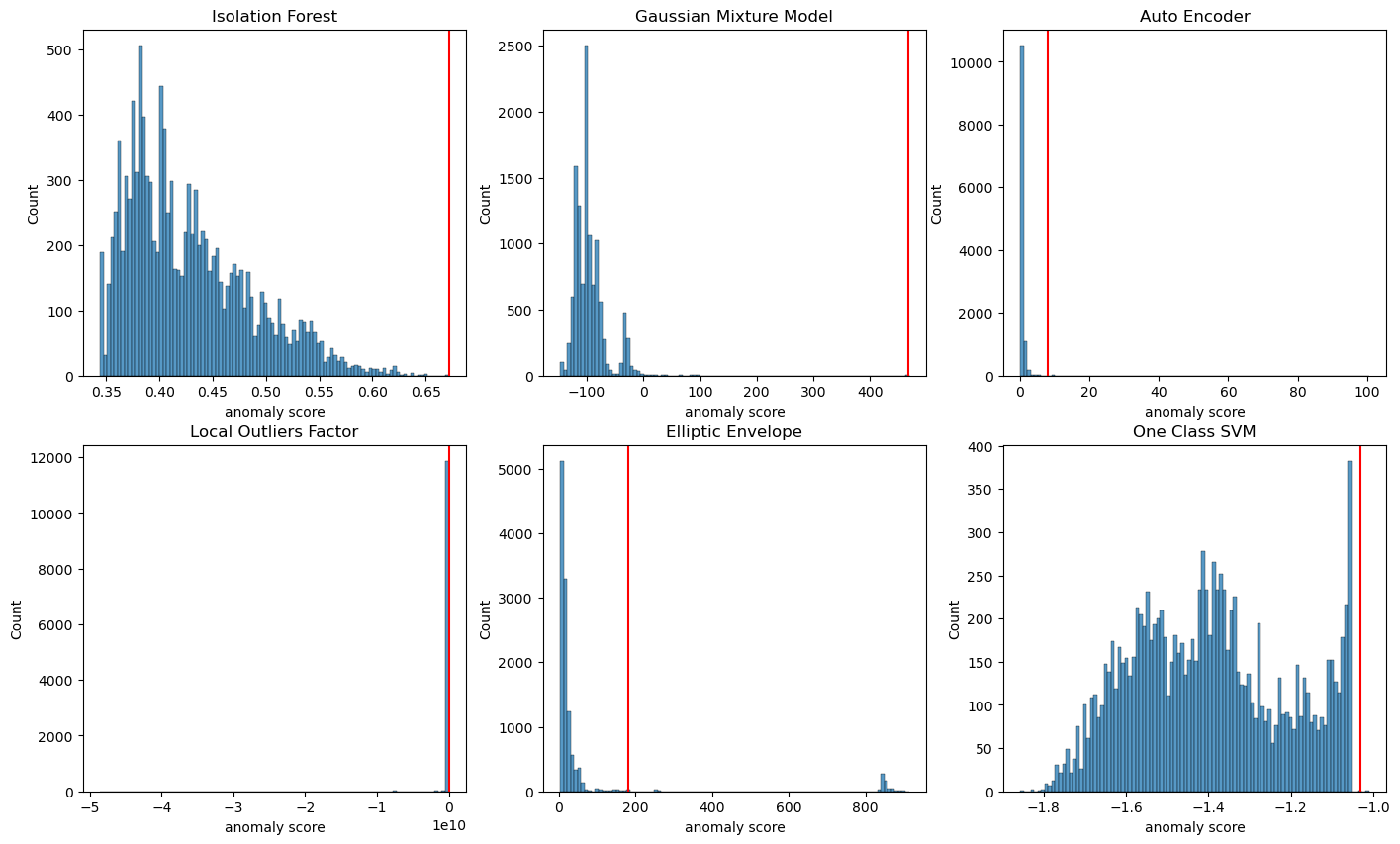}
    \caption{Comparison of all the models anomaly score distribution on the same device example. The red line represents the anomaly score of the process event identified by the experts as a malware.}
    \label{fig:model_comparison}
\end{figure}

\begin{figure}[!h]
    \centering
    \includegraphics[width=10cm]{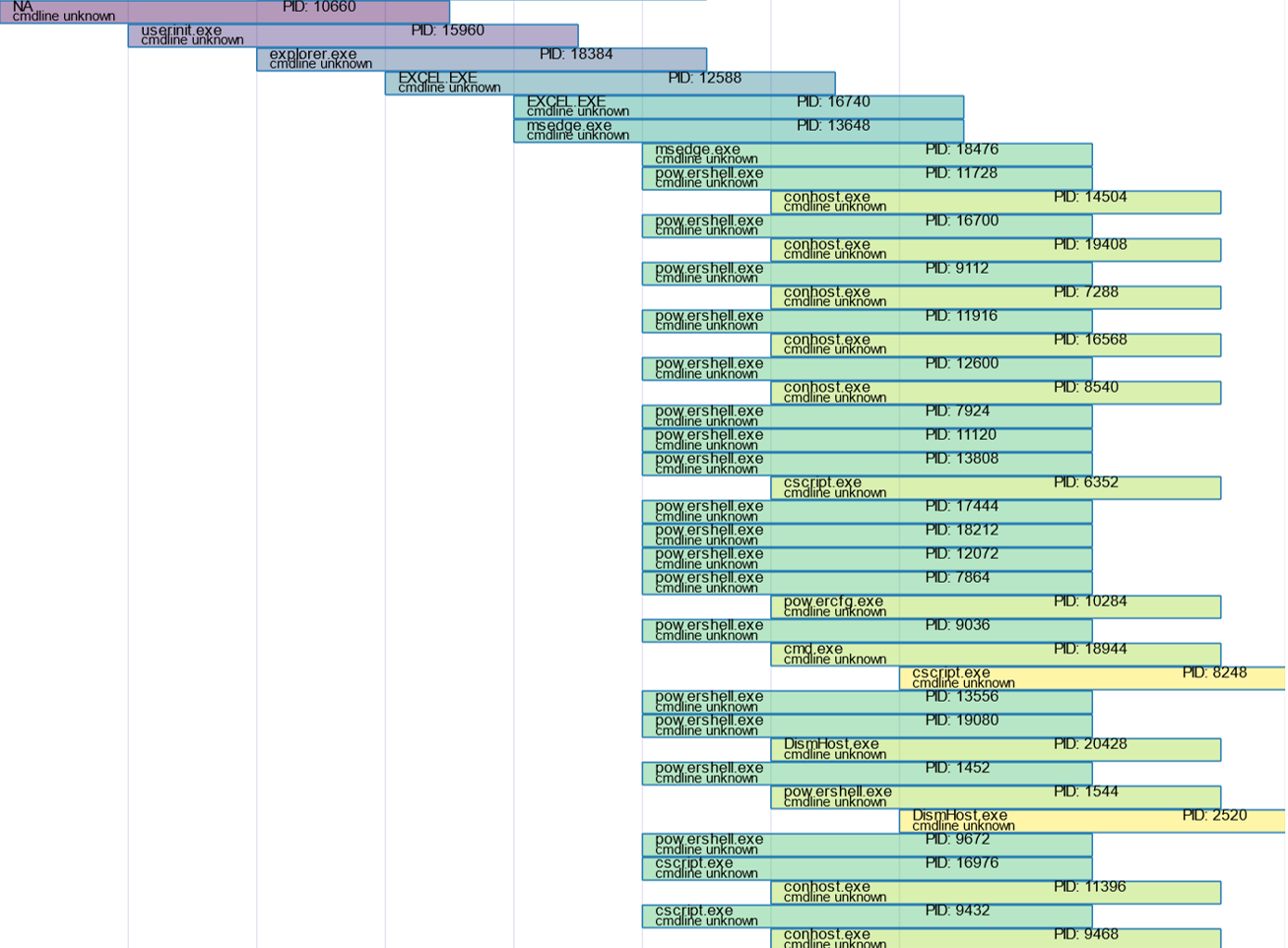}
    \caption{Example of process tree example to help the analyst identify the child, parent and sibling processes in an attack. This example is related with the attack described in Figure \ref{fig:attack_example}.}
    \label{fig:mstcipy_example}
\end{figure}

\end{document}